\documentclass[manuscript]{acmart}
\AtBeginDocument{%
  \providecommand\BibTeX{{%
    \normalfont B\kern-0.5em{\scshape i\kern-0.25em b}\kern-0.8em\TeX}}}

\setcopyright{acmcopyright}
  
\copyrightyear{2021}
\acmYear{2021}
\setcopyright{acmcopyright}\acmConference[MobileHCI '21]{23rd International Conference on Mobile Human-Computer Interaction}{September 27-October 1, 2021}{Toulouse \& Virtual, France}
\acmBooktitle{23rd International Conference on Mobile Human-Computer Interaction (MobileHCI '21), September 27-October 1, 2021, Toulouse \& Virtual, France}
\acmPrice{15.00}
\acmDOI{10.1145/3447526.3472060}
\acmISBN{978-1-4503-8328-8/21/09}
\begin{document}

\title{ThumbTrak: Recognizing Micro-finger Poses Using a Ring with Proximity Sensing}

\author{Wei Sun}
\affiliation{%
    \institution{Cornell University}
    \country{United States}
    }
\affiliation{%
    \institution{Institute of Software, Chinese Academy of Sciences}
    \country{China}
}
\additionalaffiliation{%
  \institution{School of Computer Science and Technology, University of Chinese Academy of Sciences, China}
}

\author{Franklin Mingzhe Li}
\affiliation{%
    \institution{Carnegie Mellon University}
    \country{United States}
 }

\author{Congshu Huang}
\affiliation{
    \institution{Cornell University}
    \country{United States}
}
\affiliation{
    \institution{Peking University}
    \country{China}
}

\author{Zhenyu Lei}
\affiliation{
    \institution{Cornell University}
    \country{United States}
}
\affiliation{
    \institution{Huazhong University of Science and Technology}
    \country{China}
}

\author{Benjamin Steeper}
\affiliation{%
    \institution{Cornell University}
    \country{United States}
 }

\author{Songyun Tao}
\affiliation{
    \institution{Cornell University}
    \country{United States}
}

\author{Feng Tian}
\affiliation{%
    \institution{School of Artificial Intelligence, University of Chinese Academy of Sciences, Beijing}
    \country{China}
}
\additionalaffiliation{%
    \institution{State Key Laboratory of Computer Science, Institute of Software Chinese Academy of Sciences, China}
}
 
\author{Cheng Zhang}
\authornote{Corresponding author}
    \affiliation{%
    \institution{Cornell University}
    \country{United States}
 }

\renewcommand{\shortauthors}{Sun et al.}
\renewcommand{\shorttitle}{ThumbTrak}

\newcommand{\MarkRed}[1]{\textcolor{red}{\textbf{} #1}}

\begin{abstract}
ThumbTrak is a novel wearable input device that recognizes 12 micro-finger poses in real-time. Poses are characterized by the thumb touching each of the 12 phalanges on the hand. It uses a thumb-ring, built with a flexible printed circuit board, which hosts nine proximity sensors. Each sensor measures the distance from the thumb to various parts of the palm or other fingers. ThumbTrak uses a support-vector-machine (SVM) model to classify finger poses based on distance measurements in real-time. A user study with ten participants showed that ThumbTrak could recognize 12 micro finger poses with an average accuracy of 93.6\%. We also discuss potential opportunities and challenges in applying ThumbTrak in real-world applications.



\end{abstract} 


\maketitle

\section{Introduction}

The state-of-art interaction techniques on wearable computers (e.g., smartwatch, glass) are dominated by speech-based and touchscreen-based interaction, which have not yet provided the optimal interaction experience for certain scenarios. Speech interaction may not work well in public spaces, where environmental noise may be high or talking out loud might be socially inappropriate \cite{li2017braillesketch}. Touch-based interfaces on wearables are often limited by the screen sizes \cite{oney2013zoomboard}. Therefore, the expressiveness of the touch interaction on wearables is relatively limited, compared to the touch interaction on tablets or phones. Furthermore, touch-based interaction demands exclusive eye-contact, which may not always be convenient.
For instance, if a phone call comes in during a meeting, the user may want to immediately dismiss the phone call, instead of locating the 'reject' button on a screen or issue voice commands \cite{sun2021teethtap}. Therefore, there is a need for a discrete and eyes-free input technique to promote wearable interaction experience in these scenarios.

\begin{figure}[t]
  \includegraphics[width=0.6\linewidth]{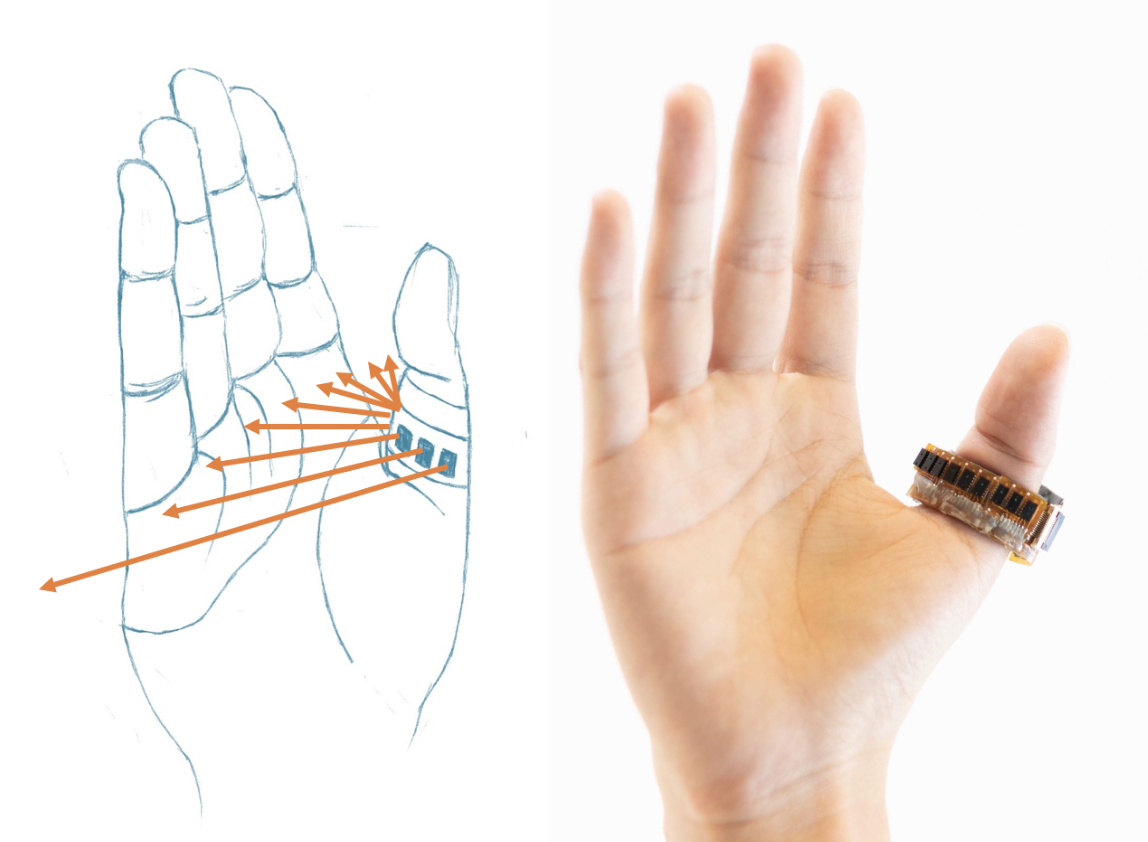}
  \caption{ThumbTrak System}
  \label{fig:Phalanges}
  \Description{This figure contains two sub-figures. The left one shows the sketch of a hand wearing a Thumbtrak system and nine red arrows that show the position and the orientation of each sensor. The right one shows a photo that a hand wearing a Thumbtrak system.}
\end{figure}

Existing work has explored various approaches to enable hand or finger gestures for wearable devices (e.g., \cite{mujibiya2013sound,kim2012digits}). Many of them rely on wearing sensor units on the wrist or forearm to recognize finger or hand gestures (e.g., \cite{kim2012digits,gong2017pyro,zhang2015tomo}). However, such methods either require heavy instrumentation, which might be obtrusive to the user, or the recognition performance from sensor units on the wrist or forearm might be affected in certain situations, such as the occlusions from clothing or certain hand angles. Therefore, we are interested in exploring ring-based approaches because many people are used to wearing rings in everyday activities and it has potentials to detect various hand/finger gestures with less constraints (e.g., hand angles or clothing conditions) than prior approaches.

In our work, we present ThumbTrak (as shown in Figure \ref{fig:Phalanges}), a thumb-ring based eyes-free pose input technique using proximity sensing. The thumb-ring is built with a flexible printed circuit board hosting nine proximity sensors, which measure distances from the thumb to various parts of the hand. Because these distances vary depending on the user's hand position, ThumbTrak can distinguish which finger phalanges of the 12 phalanges the thumb touches in real-time using a support-vector-machine (SVM) model based on measured distances. The layout of the 12 finger poses (three phalanges per finger $\times$ four fingers) is similar to a T9 keyboard, which can be naturally extended to input numbers. A user study with ten participants showed that ThumbTrak could recognize these 12 micro finger poses with an average accuracy of 93.6\%. 

The contributions of our paper are:
\begin{itemize}
    \item A novel thumb-ring-based input method which recognizes 12 micro-finger poses in real-time using proximity sensing and machine learning.
    \item A user study with 10 participants which evaluated the performance of the system (over 93\%) and presented analysis on sensor numbers and positions.
    \item A discussion of opportunities and challenges about how to apply this technology in a wider range of real-world applications. 
\end{itemize}

\section{Related Work}
Gesture recognition has always been a classic research topic in the HCI community. In this section, we introduce and discuss prior work in wearable gesture input techniques involving one-handed input methods and ring-based approaches.

\subsection{One-handed Gesture Input}

Naturally, one-handed input allows users to input information with just one hand - often a more convenient option compared to interactions requiring both hands. A variety of one-handed input technologies have been developed. Since armbands and wristbands are commonly worn accessories in society, many prior works use these two form factors to detect discrete hand or finger gestures. A variety of different sensing modalities have been employed to detect these gestures, including electric impedance tomography \cite{zhang2015tomo,zhang2016advancing}, pressure sensing \cite{dementyev2014wristflex}, acoustic sensing \cite{zhang2018fingerping,deyle2007hambone,laput2016viband,harrison2010skinput,kubo2019audiotouch}, Electromyography (EMG) \cite{lu2014hand,saponas2009enabling}, magnetic sensing \cite{huang2016digitspace}, motion sensing \cite{lu2014hand,wen2016serendipity,xu2015finger}, electrical sensing \cite{zhang2016skintrack}, cameras \cite{loclair2010pinchwatch,way2014usability,hu2020fingertrak} and proximity sensing \cite{fukui2011hand,gong2016wristwhirl}. For example, Lu et al. \cite{lu2014hand} leveraged four surface electromyography sensors to recognize hand gestures such as grasping or opening the hand. However, prior research either required heavy instrumentation or mostly focused on recognizing mid- or large-scale one-hand gestures (e.g., hand movements, hand grasping), which lacks explorations on micro-finger one-handed poses (e.g., distinguishing which finger phalanges of the 12 phalanges the thumb touches).

\subsection{Gesture Input with a Ring}
Rings are another accessory that people are accustomed to wearing. Since they sit on the fingers (or thumb) themselves, they are arguably a more convenient method for sensing micro finger poses compared to wrist- or arm-related form factors. Consequently, many previous ring-based input techniques have been invented, recognizing discrete finger or hand gestures using a wide variety of sensing modalities. Previously explored modalities have included visual sensing (cameras) \cite{chan2015cyclopsring}, motion sensing \cite{zhang2017fingersound,zhang2017fingorbits,gupta2019rotoswype}, acoustic sensing \cite{zhang2018fingerping}, magnetic sensing \cite{ashbrook2011nenya,chen2013utrack}, and infrared reflection sensing \cite{ogata2012iring}. Ring-based input projects have also been invented to track continuous finger input on 2D surfaces \cite{kienzle2014lightring} and thumb input in 3D space \cite{chen2013utrack}. 

\begin{figure}[t]
  \centering
  \includegraphics[width=0.4\columnwidth]{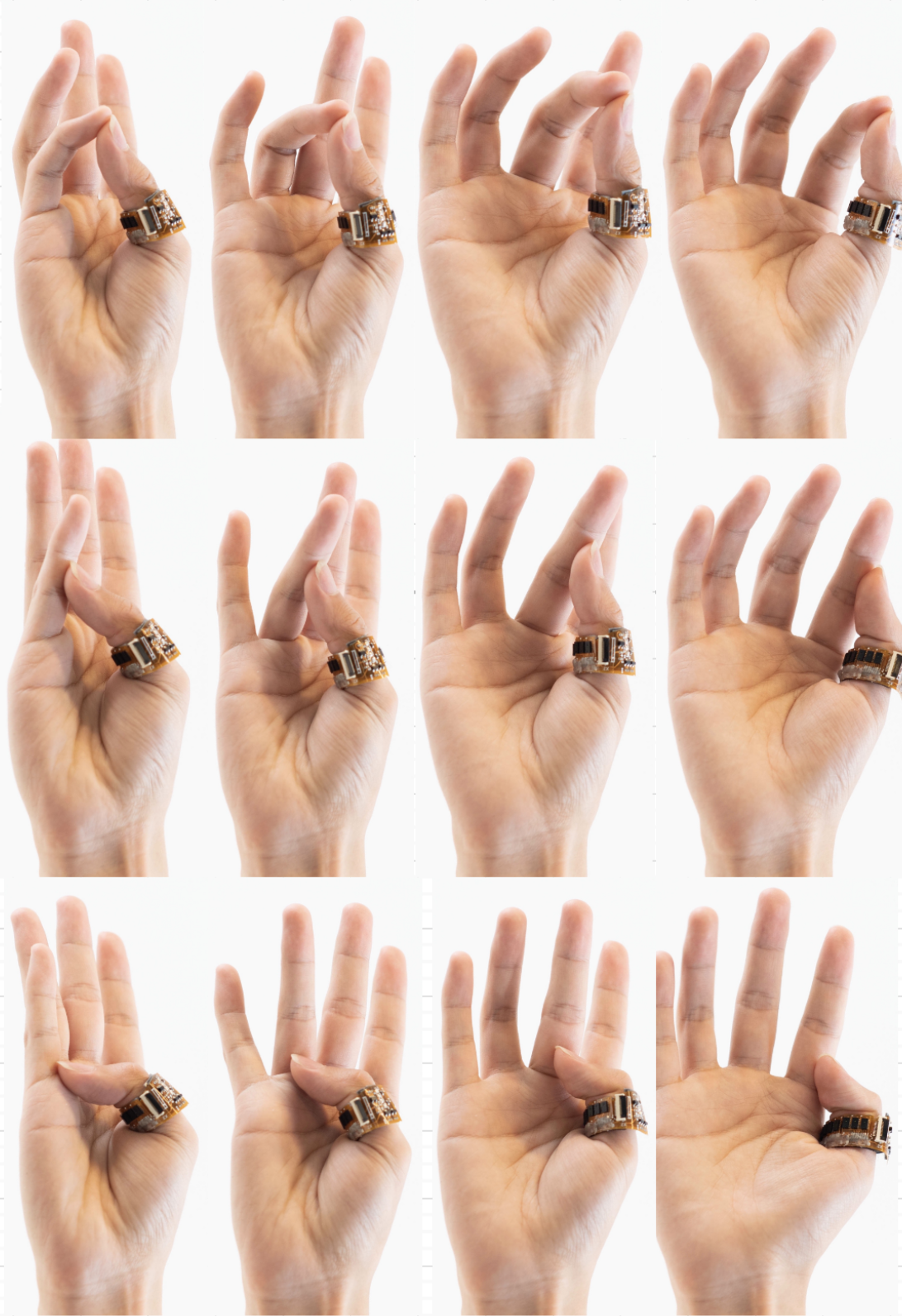}
  \caption{ThumbTrak Pose Family}
  \label{fig:PoseFamily}
  \Description{This figure contains twelve hand poses with corresponding pictures in three rows. The first row: thumb touching little finger distal, ring finger distal, middle finger distal, and index finger distal. The second row: thumb touching little finger middle, ring finger middle, middle finger middle, and index finger middle. The third row: thumb touching little finger proximal, ring finger proximal, middle finger proximal, and index finger proximal.}
\end{figure}

The most similar previous work to ours is FingerPing \cite{zhang2018fingerping}, which also recognizes thumb touches on the 12 finger phalanges. The system recognizes these poses via active acoustic sensing with a thumb-ring, and a wristband \cite{zhang2018fingerping}. Compared to FingerPing, ThumbTrak only requires one piece of form factor (the thumb ring), which is arguably more practical, since it requires less instrumentation. Furthermore, ThumbTrak is potentially more comfortable comparing to FingerPing \cite{zhang2018fingerping}, given there is no vibration on the thumb. More importantly, FingerPing has a low cross-section performance, due to inconsistencies in ring and wristband placement, which results in inconsistent signals causing acoustic impedance to vary. Alternatively, ThumbTrak uses proximity sensing - a relatively more robust cross-session solution. We present a promising cross-session recognition performance report later on in the paper.

\section{Design and Implementation of ThumbTrak}
\subsection{Pose Design}
Inspired by the natural similarity of the layouts between the finger phalanges and the T9 keyboard \cite{kubo2019audiotouch}, we designed 12 micro-finger poses (as shown by Figure \ref{fig:PoseFamily}), involving thumb touching on each of the three phalanges of the index, middle, ring, and little fingers. These micro-gestures have fewer movement distinctions comparing with other hand or finger gestures (e.g., fist and thumbs-up) and they allow input in a more discreet fashion for privacy. Furthermore, adopting an existing input layout is more straightforward to map the input poses to different functionalities in applications. For instance, the 12 poses construct a three (phalanges) by four (fingers) matrix, which can be used to input numbers. A subset of these 12 micro-finger poses can also be used as a directional pad.



\subsection{Hardware Design}
\begin{figure}[t]
\centering
  \includegraphics[width=0.7\columnwidth]{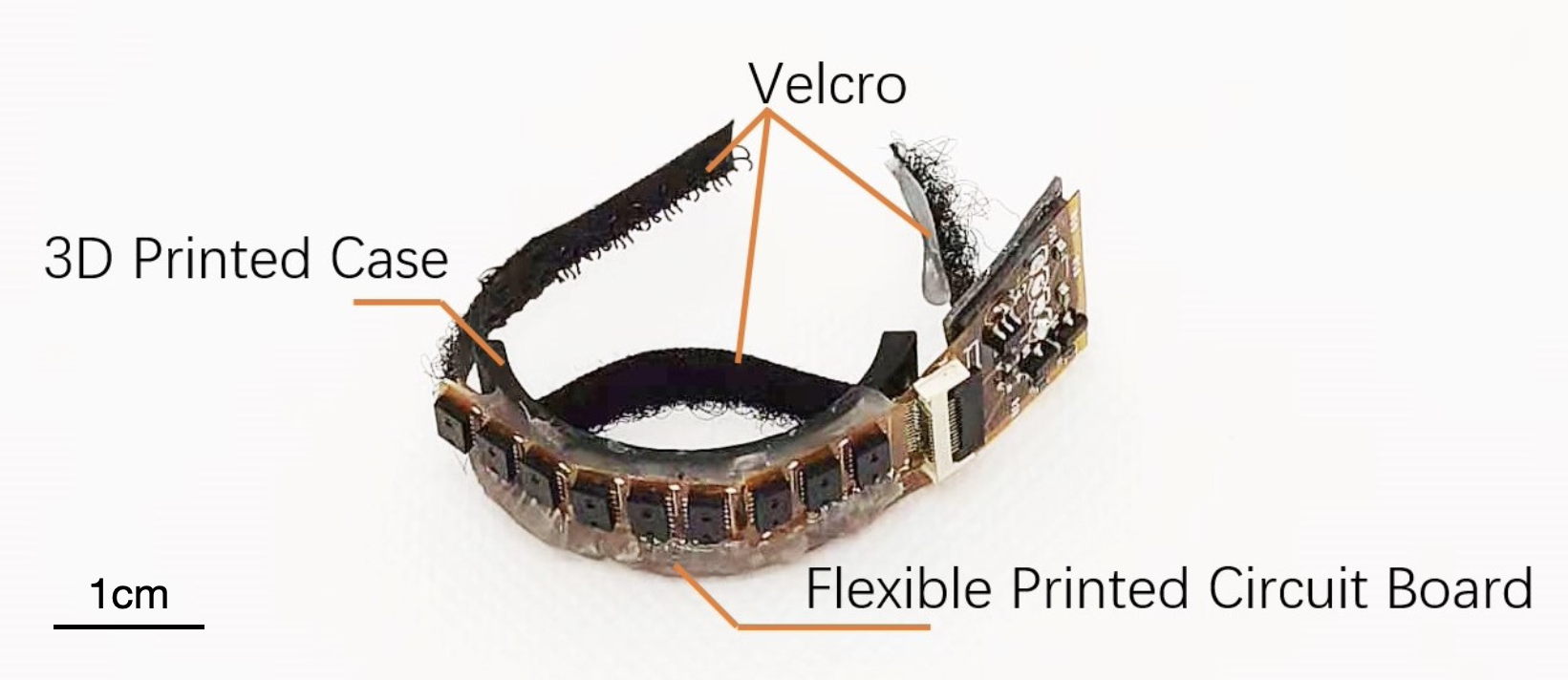}
  \caption{ThumbTrak Hardware}
  \label{fig:hardware1}
  \Description{This figure shows the thumbtrak from an overview. It combines a 3D printed case, velcro, and a flexible printed circuit board.}
\end{figure}

Figure \ref{fig:hardware1} illustrates ThumbTrak's ring, which comprises three major components: 1) a flexible printed circuit board (FPCB), 2) a 3D printed case to provide support, and 3) a Velcro band. The FPCB board hosts nine proximity sensors (VL6180X)\footnote{\url{https://www.st.com/resource/en/datasheet/vl6180x.pdf}}, and they operate at around 10Hz. We choose FPCB over a traditional printed circuit board (PCB) because FPCB is bendable and flexible, allowing it to wrap around the thumb to better measure distances in multiple directions. The more directions we can measure distances from, the more informative the sensor readings will provide. The FPCB is glued on a 3D printed case for more robust support. We use a Velcro band to fit the ring tightly and comfortably around thumbs of different sizes. The sensor readings are communicated from FPCB to a Teensy 3.2 using wired $\mathrm{I^2C}$ communication, which eventually sends the data to a python program running on a MacBook Pro for online classification via serial communication. We sequentially activate and change each proximity sensor's address to enable $\mathrm{I^2C}$ communication. In the device testing, we found that ThumbTrak consumes around 120mW power.

\subsection{Data Processing Pipeline}

We developed a Python program to continuously capture and analyze the stream of sensor data for real-time pose recognition. From our internal testings between researchers, we found that the distance between the thumb and any given part of the palm did not exceed 150mm when performing candidate poses. If a distance value does exceed 150mm, we set its value to 150mm to prevent erroneous sensing results due to the out of range sensing distances, which signifies the sensor is out of range. Many out of range sensors are an indicator of ``no-pose'' status.

\begin{figure}[t]
  \centerline{
  \includegraphics[width=0.8\columnwidth]{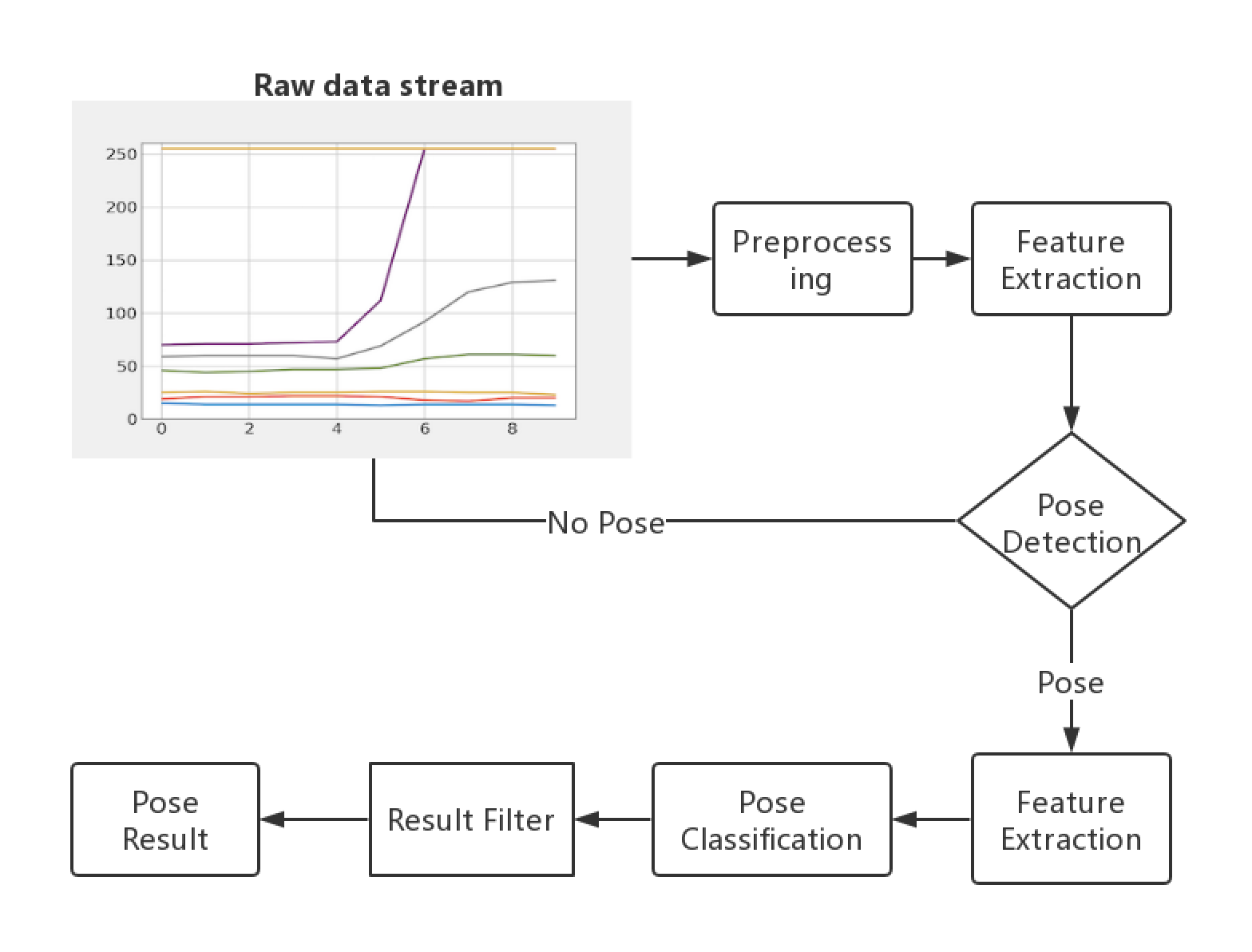}}
  \caption{Data Processing Pipeline}
  \label{fig:pipeline}
  \Description{This figure shows the data processing pipeline from raw data to extract features to determine whether it is a pose or not. Then it goes to use pose features to classified pose, and then use a filter to output pose results.}
\end{figure}


Every sample iteration, we gather each sensor's distance value for a total of nine initial features. We then use these nine values to derive two extra features: the number (quantity) of out of range sensors and the average value of all in-range sensors. This makes for a total of 11 features. We then feed all 11 features into a support vector machine model (SVM classifier with a linear kernel) to predict whether or not a pose is being performed. If the SVM model, which is responsible for segmentation, predicts a pose is being performed, we pass that segment values on to a second SVM model (with linear kernel) for classification. This second SVM classifier is responsible for predicting which finger pose of the 12 poses was performed.

To prevent potential system noises, we leveraged majority voting to detect the poses in real-time. For each data sample, we extract features, which are sent to the first SVM model to estimate whether it is a target pose or noise (non-pose). If the first SVM recognizes a data sample as a gesture, we activate the second SVM and classify the pose as one of the 12 micro-finger poses. At a specific time, if the previous 15 data points (1.5 seconds) are classified as poses, we then perform the majority voting. As the thumb may not be stabilized in the first half second, we only consider the classification result from the later one second for majority voting. Then we output the labels that represent the majority of the later 10 data points of the entire 15 data points as the gesture recognition result. Furthermore, if the SVM model outputs five consecutive results as noise (non-pose), ThumbTrak recognizes this as a reset for micro-finger pose recognition.



\subsection{Calibration}
Remounting the ring can introduce displacements on the sensor positions, which can impact the performance of the system. To address this issue, we introduce a calibration method in our system to help the users to remount the ring to its original position with minimal effort. We determined proximity to its original position by outputting the absolute value of the difference between its current distance and its original distance for two poses: little-proximal and little-distal (the furthest fingers from the thumb). In the calibration process, we first asked participants to perform the two poses (i.e., little-proximal and little-distal). The researcher then empirically and iteratively adjust the ring position based on the output value. In our user evaluation, we set the average offset threshold value per sensor as 0.7mm.



\section{User Study}
To evaluate ThumbTrak, we conducted a user study with ten participants with an average age of 22 (three female). All participants were right-handed. The user study comprises three main phases: 1) practicing phase---participants get to know the micro-finger poses and our prototype; 2) training phase---researcher collects training data with participants to generate our recognition model; 3) testing phase---participants are asked to perform different micro-finger poses to evaluate the performance of our system. Before conducting the user study with ten participants, we piloted our system and study process with two other participants. The study took around 60 minutes on average. The whole study procedure was approved by the institutional review board (IRB).

\subsection{Practicing Phase}
At the beginning of the user study, a researcher introduced our project and outlined the study's procedure. The researcher then helped the participant to fasten the ring, ensuring it be fixed tightly yet comfortably around the user's thumb to avoid unwanted rotation. Next, the researcher demonstrated how to perform each pose, instructing them to follow along with their dominant hand. After this, the participant familiarizes herself with different micro-finger poses. Participants were seated for the entirety of the experiment.

\subsection{Training Phase}
To collect training data, participants were asked to participate in three training sessions. In each training session, the participant was asked to perform five instances of each of the 13 poses (including ``no-pose''). Poses were prompted by a screen displaying an image of an open hand with a red dot resting over the region of the finger to be touched. Images appeared for a total of four seconds: three seconds for participants to react and perform the pose and one second for the system to record the sensor data. All in all, each participant performed 65 randomly ordered poses continuously for each of the three training sessions. Between each training session, participants got a short break.

\subsection{Testing Phase}
Each testing session consisted of 60 poses (i.e., 12 micro-finger poses x 5). Additionally, after each completed pose, the real-time result was presented on the screen. If the result matched the requested pose, the interface turned green. Otherwise, the interface turned blue and displayed the falsely predicted pose. Every time when participants received the results, they were asked to reset the micro-finger pose to no-pose for segmentation purposes. The testing phase contains three sessions: we called the first two sessions as in-session tests, which focus on the overall performance of the system, and the final one as the cross-session test, which mainly tests for the effects of remounting of the device. Between each testing session, participants got a short break. After finishing the last in-session test, a researcher helped the user to remove the ring. After a short rest, the researcher affixed the ring back onto the user's thumb and proceeded to undergo the calibration process. After calibration, the participant continued with the cross-session study, which was identical to in-session tests.

After completing all sessions, each participant was asked to fill out a questionnaire collecting some personal information regarding age, gender, hand dominance, wearable-device use, and thumb size. Participants also provided subjective feedback on the device, the comfort level, and perceived social appropriateness. Lastly, each participant received \$10 for participating in the study.

\section{Results}
Throughout the entire user study, all participants correctly performed every requested hand pose as the stimuli indicated on the screen, both in training and testing sessions. In total, across all 10 participants, we collected 1950 (13 poses x 150) instances in training sessions and 1800 (12 poses x 150) instances in testing sessions. Of the 1800 total testing instances, 1200 were in-session, and 600 were cross-session. If the recognition results in real-time matched the stimuli poses on the screen, we recorded it as a correct instance. Otherwise, it was a classification. The results reported in this section were the real-time performance recorded during the user study.

\subsection{In-session Testing}

\begin{figure}[b!]
\centering
\includegraphics[width=0.8\columnwidth]{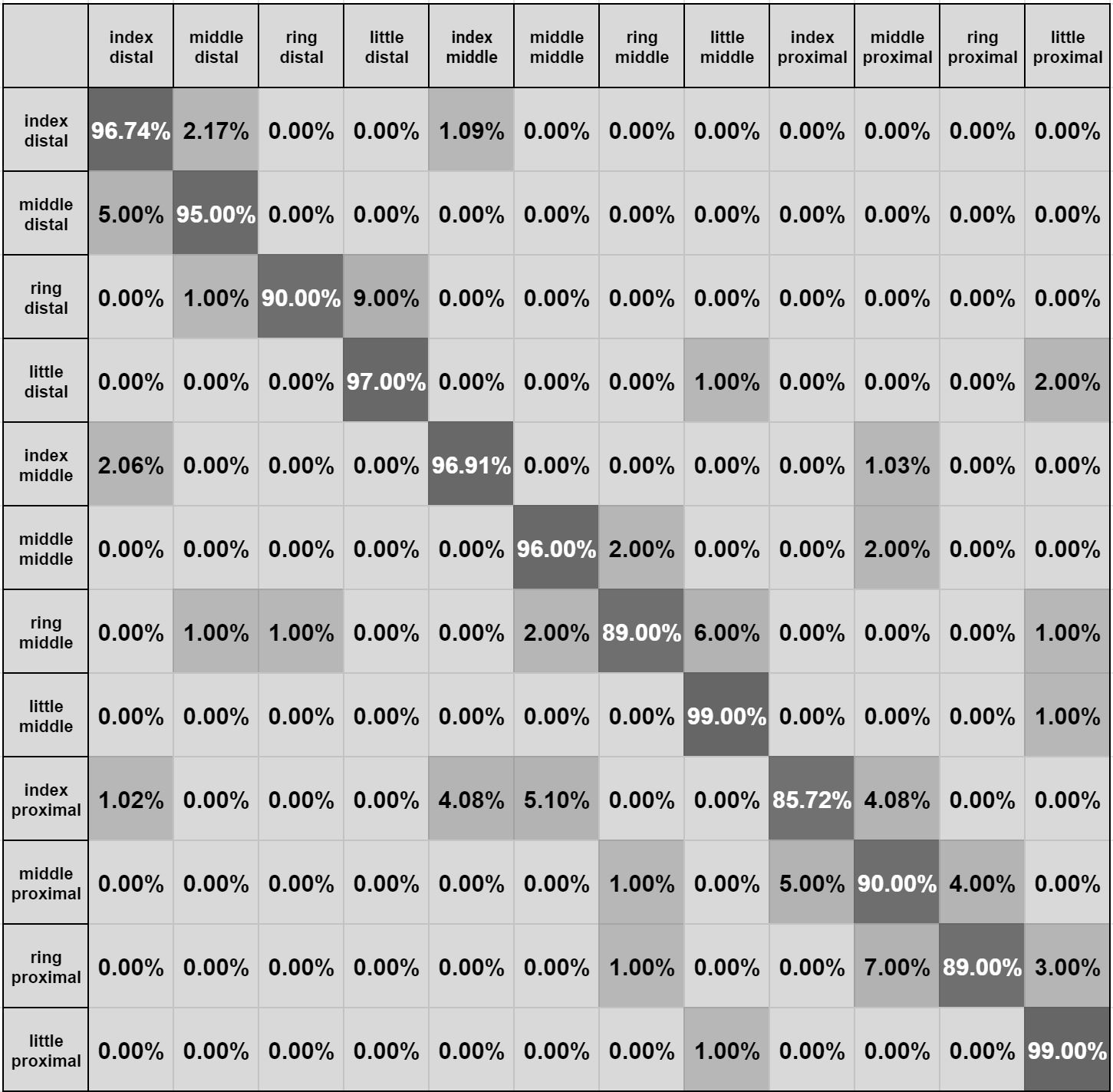}
\caption{Confusion Matrix for recognizing poses for in-session testing}
\label{fig:cm_in_session}
\Description{This figure shows a confusion matrix with rows are true labels and columns are predicted labels. For each value in the confusion matrix, the background color close to black if the value is close to 100 and the background color close to white if the value is close to 0. It shows a black diagonal pattern from top left to bottom right.}
\end{figure}
Among the performance of ten participants, the average accuracy across all poses was 93.6\%. The confusion matrix in Figure \ref{fig:cm_in_session} shows the average accuracy results of all participants for each hand pose. Recognition was most accurate for poses involving the little finger and least accurate at the index proximal position (Figure \ref{fig:cm_in_session}). There is a common confusion among the proximal positions of the index, middle, and ring fingers. Specifically, we uncovered that index proximal (Accuracy=85.72\%) was misclassified as index middle, middle middle and middle proximal (Figure \ref{fig:cm_in_session}). From analyzing these poses and feedback from participants, we found that some participants have a hard time performing the pose to touch the index proximal with thumb, because index proximal is the closest position to the thumb. Furthermore, we revealed that the ``too close'' situation made sensors have a smaller variability in distance comparing with fingertips, which results in the misclassification to index middle, middle middle, and middle proximal (Figure \ref{fig:cm_in_session}).

\begin{figure}[t]
\centering
\includegraphics[width=0.5\columnwidth]{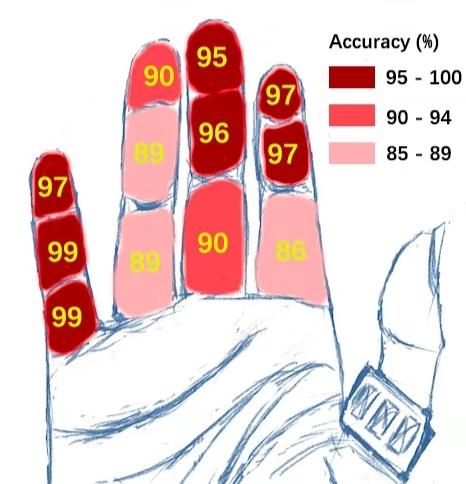}
\caption{Accuracy for 12 Phalanges (Rounded to nearest integer)}
\label{fig:accuracy_phalanges}
\Description{This figure shows the accuracy for 12 phalanges on a hand. For each value of the accuracy, the background color close to red if the value is close to 100 and the background color close to white if the value is close to 0. This figure shows the accuracy for the index (97, 97, 86, top to bottom), middle (95, 96, 90, top to bottom), ring (90, 89, 89, top to bottom), and little finger (97, 99, 99, top to bottom).}
\end{figure}

\begin{figure}[b]
  \centering
  \includegraphics[width=0.8\columnwidth]{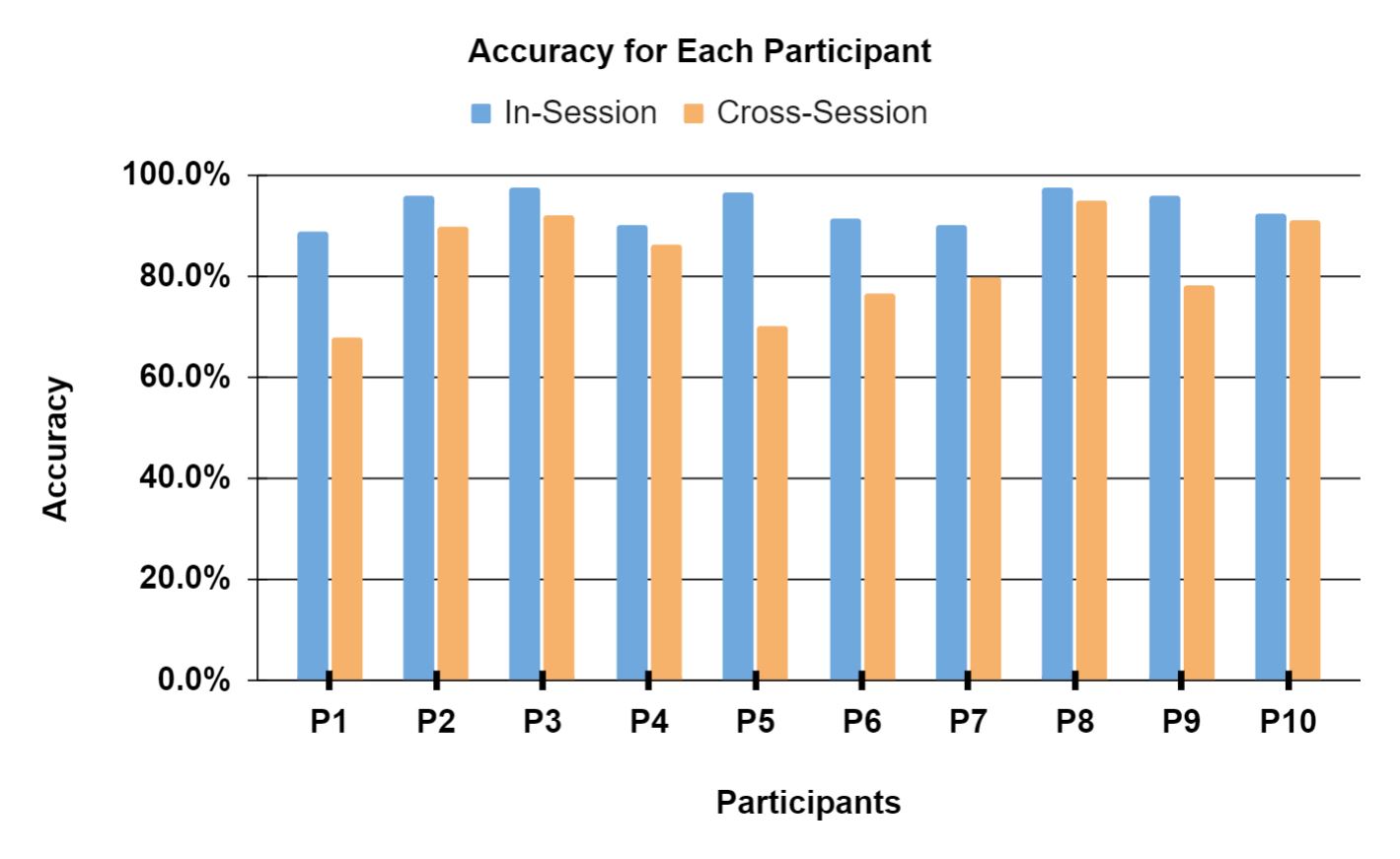}
  \caption{Accuracy for each participant}
  \label{fig:barchart_user_study}
  \Description{This figure is a bar chart of the recognition accuracy for 12 poses. X-axis: each participant. Y-axis: the recognition accuracy in percentage. For each participant, the figure shows two bars in blue (in-session) and yellow (cross-session).}
\end{figure}
We then visualized the recognition performance on different finger positions (Figure \ref{fig:accuracy_phalanges}), which depicts the accuracy for each of the 12 phalanges across all participants. Beyond the misclassification of the index proximal, we realized that the recognition performance of the ring finger is relatively worse than the middle finger and the little finger. In our study, we found that it is a natural behavior for participants to move their middle finger when they try to use their thumb to touch the ring finger, which may affect the recognition performance. In Figure \ref{fig:barchart_user_study}, we present the overall performance for each participant during the in-session tests and the cross-session test. The figure illustrates that universally, all participants were able to achieve a relatively high average accuracy (Figure \ref{fig:barchart_user_study}). Overall, we found that the highest accuracy achieved was 97.50\% while the lowest was 88.79\%. The median accuracy was 94.10\%.

\subsection{Cross-session Testing}
In the cross-session testing, we found that the average accuracy across all participants was 82.7\%. Cross-session confusion error distributions were quite similar to in-session errors. For example, we found that index proximal was also misclassified as index middle, middle middle, and middle proximal. Furthermore, we found middle middle was the second most erroneous pose and got misclassified as index distal, index middle, ring middle, and index proximal. Figure \ref{fig:barchart_user_study} shows that accuracy varies greatly among participants, from the highest 95.00\% to the lowest 67.86\%, with an 82.9\% median accuracy. From the observation in the cross-session test when the researcher tried to assist participants in remounting the device, we discovered that the existing calibration process reduced the effort of memorizing the ring position relative to hand or finger biometrics (e.g., `the ring was worn at 1cm below the thumbnail'). 


\begin{figure}[h!]
  \centering
  \includegraphics[width=1\columnwidth]{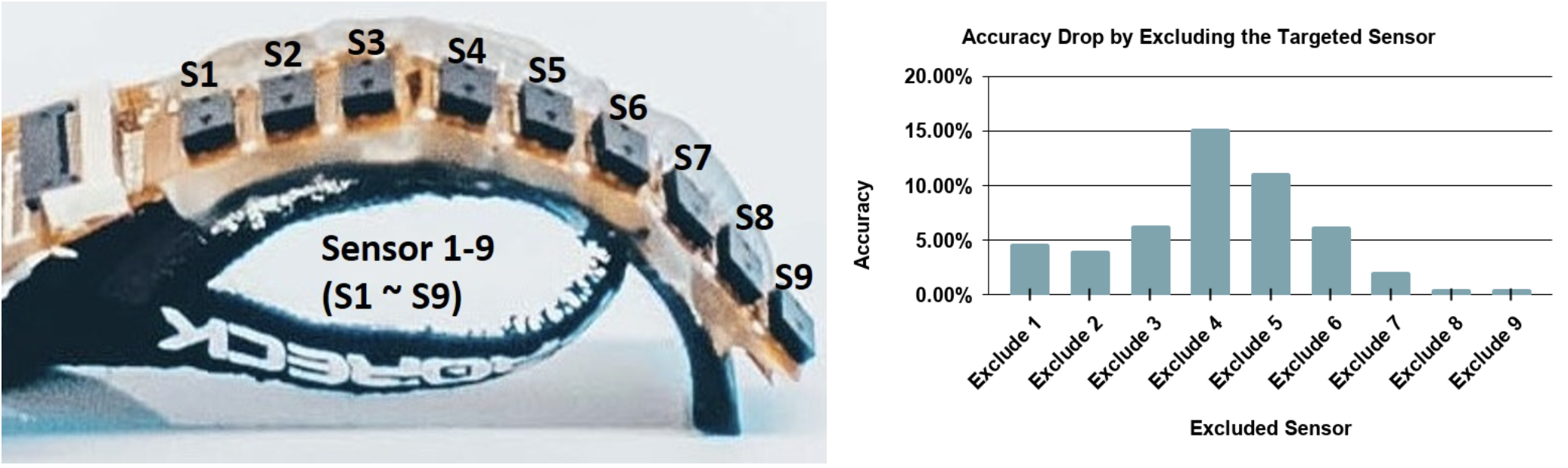}
  \caption{Left: Sensor name mapping on ThumbTrak. Right: Accuracy drop by excluding the targeted sensor}
  \label{fig:barchart_exclude_one}
  \Description{This figure contains two sub-figures. The left shows the label of nine sensors from S1 to S9. The right shows a bar chart. Y-axis is the accuracy of pose detection. The X-axis is the sensor label that excluding the targeted sensor. This right sub-figure shows that the accuracy drop start with low from excluding the sensor 1, and it keeps increasing till sensor 4 and keeps dropping till sensor 9.}
\end{figure}

\subsection{Analysis of Sensor Numbers and Positions}
In our study, we utilized nine proximity sensors on the ring to recognize micro-finger poses. We then perform further analysis on how different positions and numbers of proximity sensors may affect the recognition performance. To understand the usefulness of different sensors at different positions, we first applied the study segmentation results and removed one sensor, and used the other eight sensors for training and testing, each time we change the position of the excluded sensor. From the analysis results, we uncovered that removing sensors in the middle caused the performance to drop the most. For example, the accuracy dropped 15.25\% if we remove Sensor 4 (Figure \ref{fig:barchart_exclude_one}). On the contrary, the performance stayed around 90\% if we remove the sensor at the side of the ring.

\begin{figure}[t]
  \centering
  \includegraphics[width=0.6\columnwidth]{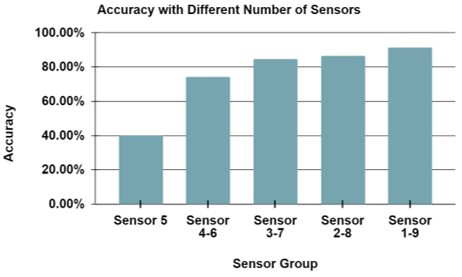}
  \caption{Accuracy with different number of sensors}
  \label{fig:barchart_accuracy_number}
  \Description{The figure shows a bar chart that the Y-axis is the accuracy of pose detection, and the X-axis is the sensor label of training sets.}
\end{figure}

To further understand the effect of using fewer sensors on the performance change, we leverage the study segmentation results and start with sensors in the middle, because we found sensors in the middle contribute to the performance the most. In this analysis, we start with Sensor 5; we then keep adding sensors surrounding the sensor group. Overall, We ended with five sensor subsets: Sensor 5, Sensor 4-6, Sensor 3-7, Sensor 2-8, and Sensor 1-9 (Figure \ref{fig:barchart_accuracy_number}). By analyzing associated performance changes (Figure \ref{fig:barchart_accuracy_number}), the accuracy stays around 40\% when we only use Sensor 5 for training and testing (Figure \ref{fig:barchart_accuracy_number}). After we evaluate the recognition performance with using Sensor 3-7, the accuracy increases significantly to 85\% (Figure \ref{fig:barchart_accuracy_number}). However, the accuracy increases less than 10\% after we add the four sensors at the side (Figure \ref{fig:barchart_accuracy_number}). We further inspect the sensor output from the proximity sensors at the side, and we find that they sometimes remain out-of-range for certain micro-finger poses. Therefore, future research could perform a similar analysis to reach a compromise between power consumption and recognition accuracy.

\subsection{Reduce Training session}
In our in-session tests, we used data from all three training sessions to train the classifiers. However, it is often advantageous for users to reduce training data collection time. To investigate this, we reduced the training data to $n$ training session(s) ($n=1,2$). This resulted in an accuracy of 83.5\% for $n=1$ and 91.2\% for $n=2$ (93.6\% for $n=3$). Thus, future developers should compromise between the effort of collecting training data and the recognition accuracy.


\section{Discussion}
In the result section, we showed the accuracy and recognition confusion in both in-session testing and cross-session testing. We then analyzed the recognition performance effects of sensor numbers and positions. We also mentioned the usefulness of calibration when remounting the device and the tradeoff of reducing training sessions. In the discussion, we will further discuss 1) applications and opportunities for ThumbTrak, 2) potential affordance of ThumbTrak, 3) pose subsets for cross-session, 4) calibration and remounting process, and 5) hardware design improvements.

\begin{figure}[t]
  \centering
  \includegraphics[width=0.6\columnwidth]{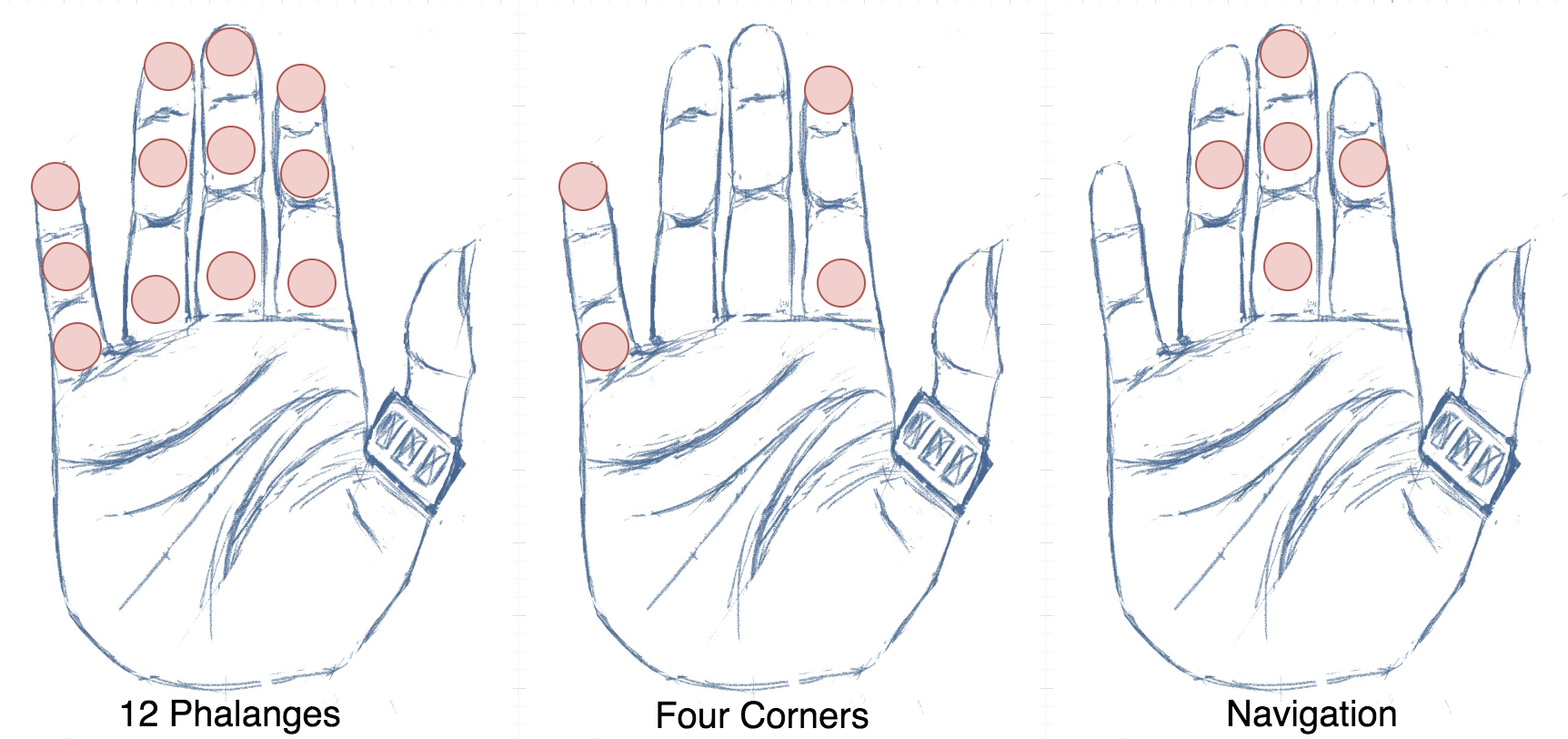}
  \caption{Layout of pose subsets}
  \label{fig:layout}
  \Description{This figure contains three sub-figures. All sub-figures show a sketch of a hand that labeled the layout of pose subsets the 12 phalanges(left), four corners(middle), and navigation(right).}
\end{figure}

\subsection{Applications and Opportunities for ThumbTrak}

\subsubsection{T9 Keyboard}
ThumbTrak offers up to 12 discrete input gestures with an average accuracy rate of 93.6\%. We found that it is natural to map these 12 micro-finger poses to a number input system due to their structural similarity (Figure \ref{fig:PoseFamily}). People could use ThumbTrak to make phone calls to their friends, and they could also utilize ThumbTrak as a T9 keyboard to accomplish text entry tasks in different messaging and social media applications. 

\subsubsection{D-Pad Control}
Certain micro-finger poses can also be chosen to form new pose subsets (Figure \ref{fig:layout}) to meet the demand of different applications. Many applications do not require all 12 poses to be functional. For instance, we can use a pose set (Figure \ref{fig:layout}) to control a D-Pad, which enables navigation through hierarchical menus for wearable devices. The buttons of the D-Pad can be naturally mapped to the index-middle, middle-distal, middle-proximal, and ring-middle positions (Figure \ref{fig:layout}). If we only include these poses in the training and testing sessions, this subset can achieve an average accuracy of 99.0\% in post-analysis (Figure \ref{fig:barchart_layout}). 

\subsubsection{Music Control}
ThumbTrak can detect the ``pose'' and the ``no-pose'' gesture. Thus, we can calculate the gesture holding time, which can provide continuous input, such as increase or decrease the music volume. The user could hold a pose to raise or lower volume and release the gesture when the volume has reached the desired level. Similarly, for music control, the user could use ThumbTrak to place commands for `Fast Forward', `Rewind', `Play', and `Pause'.

\subsubsection{Application Shortcuts}
Another interesting pose subset, the four corner phalanges (index-distal, index-proximal, little-distal, little-proximal), can also be integrated as shortcuts with many applications. For example, the user could use the index-distal pose to do mobile screen capturing when playing a game on a smartphone (e.g., PUBG MOBILE). This subset presented a 97.1\% accuracy in post-analysis (Figure \ref{fig:barchart_layout}).

\subsubsection{Virtual Reality \& Augmented Reality}
Much existing virtual reality and augmented reality systems utilize WIMP (i.e., ``Windows, Icons, Menus, Pointer'') for user interactions. Similar to prior work on enabling faster control with finger or hand gestures  \cite{kulshreshth2014exploring,bowman2001design}, ThumbTrak could also provide the opportunity of enabling fast hierarchical controls and reduce the effort of controlling the pointer movement through button and icon interactions, but without having to wear complex sensor instrumentation (e.g., sensor gloves \cite{bowman2001design}) or instrumenting a camera in front of the user \cite{fan2020eyelid,kulshreshth2014exploring}.

\subsection{Potential Affordance of ThumbTrak}
In our study, we only evaluated ThumbTrak on recognizing discrete micro-finger poses. However, in our experiment, we observed that the distance profile between thumb to other fingers could also be very informative to predict the thumb position in 3D space. One natural next step of ThumbTrak is to explore continuous tracking of the thumb position in 3D space using time-series modeling techniques, similar to the work of UTrack \cite{chen2013utrack}. This would grant ThumbTrak an even richer expressiveness in input space. Furthermore, ThumbTrak could potentially support other hand or finger gestures, such as fist and thumbs up. Because the distance between the thumb and the palm would change when performing these gestures. Therefore, future applications could explore customized finger or hand gestures to naturally and unobtrusively interact with their interfaces with ThumbTrak.

In this current research, we focused on the general proof-of-concept and simply aligned nine sensors in a line. Beyond the existing hardware setup with a 1D array of proximity sensors, an extension of ThumbTrak could change the sensor layout from 1D to 2D to enable more depth detection. The angle of each sensor, their relative position to one another, and the number of sensors all may play a critical role in improving the ring's performance. Other than proximity sensing, future research could explore how to leverage the micro depth camera on a ring-based sensing unit for hand/finger gesture recognition. Moreover, using radar sensors or acoustic sensors could also extend the potential applications of ring-based sensing units.

\subsection{Pose Subsets for Cross-session}
\begin{figure}
  \centering
  \includegraphics[width=0.6\columnwidth]{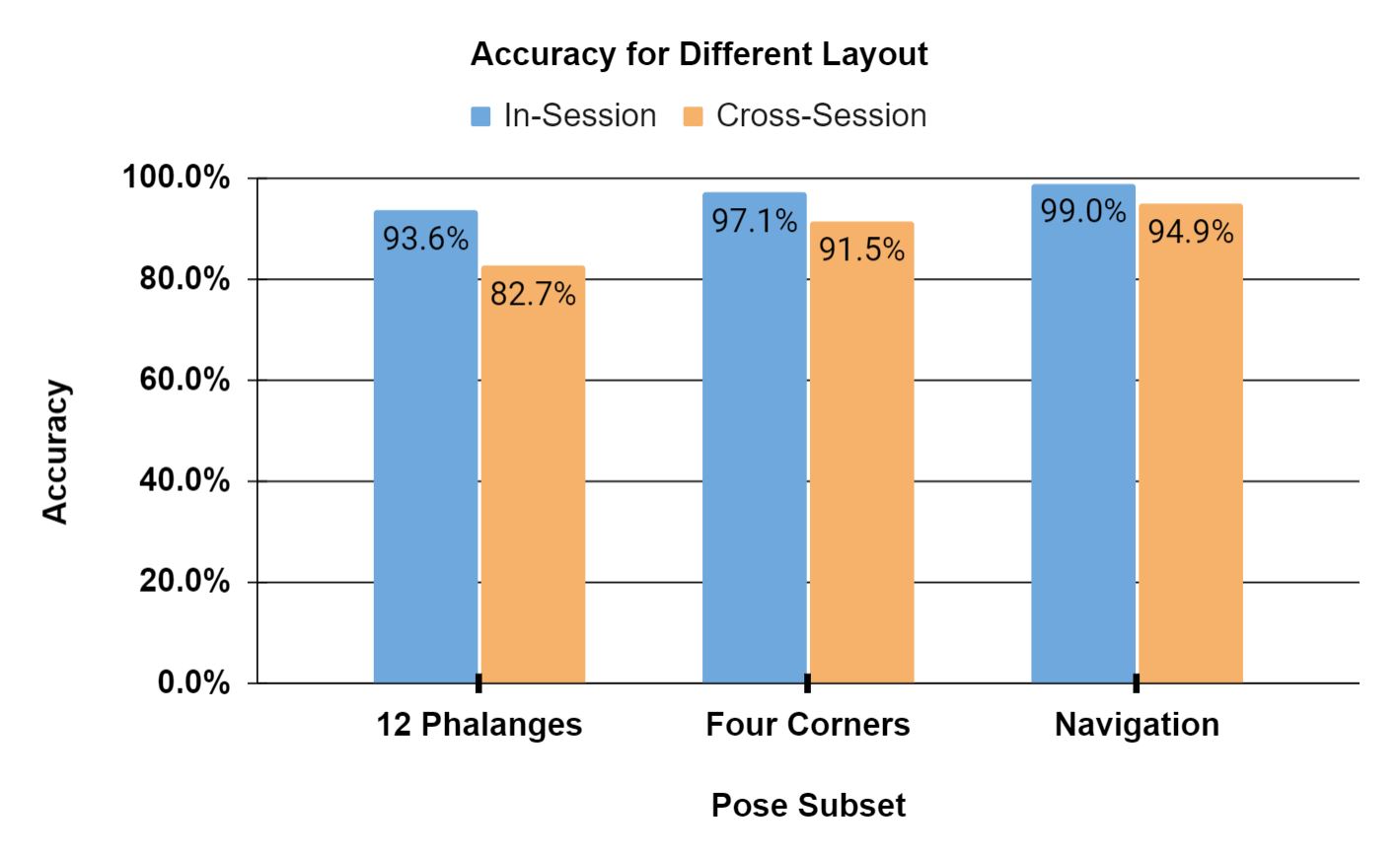}
  \caption{Accuracy for different layout of pose subsets}
  \label{fig:barchart_layout}
  \Description{This figure is a bar chart of the recognition accuracy for 3 pose subsets, the Y-axis: the percentage accuracy. X-axis: each pose subset, the 12 phalanges(left), four corners(middle), and navigation(right).}
\end{figure}

As illustrated above, some re-combined pose subsets are sufficient for multiple tasks and applications. This also applies to cross-section subsets. In our cross-session testing with 12 poses, the accuracy dropped from 93.6\% to 82.7\%. For navigation micro-finger pose subset (index-middle, middle-distal, middle-proximal, and ring-middle), we found that the accuracy only dropped 99.0\% to 94.9\% (Figure \ref{fig:barchart_layout}). In terms of the four corner phalanges micro-finger pose subset (index-distal, index-proximal, little-distal, little-proximal), the accuracy dropped from 97.1\% to 91.5\% (Figure \ref{fig:barchart_layout}). Thus, we found that using a subset of poses could enable the system to perform at high accuracy even if the participant need to remount the device, without applying any further calibration procedure.

\subsection{Calibration and Remounting Process}
In our calibration process, our algorithm design relies on the ring's position to remain stable. Once the ring shifts positions, the system may begin to misclassify hand poses. In our cross-session results, we found that the performance dropped around 10\% after remounting the device. To better assist the calibration procedure, we may include more training data from different ring positions during calibration. This may compensate for the minor displacement of the ring and provide a more adaptive system to remounting situations.

Furthermore, we found that providing calibration feedback or adjustment recommendations is important for users while remounting their devices. To provide calibration feedback, we realized that it might be useful to attach small vibration motors to different positions of the ring to provide haptic feedback to users. Furthermore, the calibration process we used in the study---just by reporting the absolute differences of the sensor values comparing with the original position---may not provide enough information to a user where she should adjust the ring position. Therefore, having a visualization tool would be beneficial to users while they are using ThumbTrak in real-life scenarios.

\subsection{Hardware Design Improvement}
In order to ensure that all participants could wear our ring, we could not simply print one ring of a standard size. So we decided to create an adjustable ring. We achieved this by 3D printing a curved frame with a hole in each end. We then fed Velcro straps through both holes so that when the strap is tightened, the curved 3D printed frame will flex and tighten around any sized thumb (as Figure \ref{fig:hardware1} shows). An issue with this design, however, is its tendency to shift positions while users perform hand poses, reducing ThumbTrak's recognition accuracy as a consequence. In addition, this hard plastic PLA may not be comfortable enough for everyday uses. A potential solution could be to replace the plastic with other more elastic materials like rubber. Increased elasticity would result in a softer, more adjustable ring. Moreover, rubber would increase friction between the ring and the skin, helping the ring to remain in a fixed position. Also, it is worth noting that although it was important to make a one-size-fits-all model for the purposes of our experiment, rings by nature are not typically designed to be one-size-fits-all. In future form factors, ThumbTrak could therefore be integrated into customizable ring sizes in the same way that rings are custom-fitted in the jewelry industry today.

\section{Limitation and Future Work}
In our paper, we conducted a quantitative study to asked participants to evaluate ThumbTrak in relatively static manners under a lab setting. However, people may leverage micro-finger poses while standing, walking, jogging, holding objects in their hands, or in complex environments, all of which were not included in the current study. Moreover, our current prototype requires wire connections to collect data for gesture recognition. In future work, we plan to first create a self-contained prototype that is self-powered and has wireless communications. We will further conduct scenario-based studies (e.g., \cite{li2019fmt,kianpisheh2019face}) to both quantitatively and qualitatively evaluate the performance of ThumbTrack and identify potential problems under different contexts.

Another limitation is that we only included the ``no-pose'' gesture as participants' hands stay still in static situations. In real-life scenarios, people may accidentally move their fingers, which may result in different sensor outputs and unintended gestures. In our evaluation, we only studied discrete gestures under a lab setting, people might use micro-finger gestures in a continuous manner. Future work should explore methods to reduce false positives(e.g., activation gesture) and enable continuous gestures for real-life scenarios. 

Furthermore, our researcher helped participants to calibrate the ring position in the cross-session testing, which limit our explorations on the behaviors of how participants remount the ring by themselves, which they will eventually do under real-life scenarios. In future work, we will conduct studies to specifically understand users' behaviors of calibration and remounting the device. Finally, our participants are mostly younger adults (average age of 22). In future work, we will include participants from different age groups. Although we understand our work could be strengthened with the above modification, we think our work---a novel thumb-ring-based input method that recognizes 12 micro-finger poses in real-time using proximity sensing and machine learning---makes novel contributions to HCI and mobile-sensing research community.

\section{Conclusion}
In this paper, we present ThumbTrak, a novel Thumb-ring based input technique able to recognize micro-finger poses in real-time with proximity sensing. A user study with 10 participants shows that ThumbTrak can recognize thumb touches on all 12 finger phalanges with an average accuracy of 93.6\%. We compared our in-session testing, and cross-session testing results and uncovered design guidelines of micro-finger pose recognition system with proximity sensing. We further discussed the potential applications for ThumbTrak (e.g., D-Pad), the accuracy of pose subsets for cross-session testing, calibration and remounting process, continuous tracking of thumb, and potential improvements of hardware design. We believe our findings would shed light on future research with micro-finger poses sensing.

\begin{acks}
This project is supported by the Department of Information Science at Cornell University. We appreciate the constructive reviews from reviewers and early discussion and feedback from colleagues in the Smart Computer Interfaces for Future Interactions (SciFi) Lab at Cornell University.

\end{acks}

\bibliographystyle{ACM-Reference-Format}
\bibliography{main}

\end{document}